\documentclass[%
reprint,
superscriptaddress,
showpacs,
 amsmath,amssymb,
 aps,
prd,
]{revtex4-1}

\usepackage{graphicx}
\usepackage{dcolumn}
\usepackage{bm}
\usepackage{amsmath}
\usepackage{autobreak}
\usepackage{hyperref}
\usepackage[mathlines]{lineno}
\usepackage{natbib}
\usepackage{multirow}
\usepackage{booktabs}
\usepackage{xspace}
\usepackage{float}
\usepackage{diagbox}
\usepackage{subfigure}
\usepackage{makecell}

\begin{document}

\preprint{APS/123-QED}

\title{A three-step proposal for searching for light shining through walls\\in the X-ray band at the High Energy Photon Source}

\author{R.~T.~Chang}
\affiliation{Key Laboratory of Beam Technology of Ministry of Education, School of Physics and Astronomy, Beijing Normal University, Beijing 100875, China}

\author{S.~Feng}
\affiliation{School of Nuclear Science and Technology, University of South China, Hengyang 421001, China}

\author{X.~P.~Geng}
\affiliation{Institute of Applied Physics and Computational Mathematics, Beijing 100094, China}

\author{C.~Y.~Hu}
\affiliation{School of Nuclear Science and Technology, University of South China, Hengyang 421001, China}

\author{H.~T.~Hu}
\affiliation{Key Laboratory of Beam Technology of Ministry of Education, School of Physics and Astronomy, Beijing Normal University, Beijing 100875, China}

\author{Y.~Y.~Hu}
\affiliation{Key Laboratory of Particle and Radiation Imaging (Ministry of Education) and Department of Engineering Physics, Tsinghua University, Beijing 100084, China}
\affiliation{NUCTECH Company, Beijing 100084, China}

\author{Y.~H.~Huang}
\affiliation{School of Nuclear Science and Technology, University of South China, Hengyang 421001, China}

\author{B.~Liao}
\affiliation{Beijing Normal University at Zhuhai, Zhuhai 519087, China}
\affiliation{Key Laboratory of Beam Technology of Ministry of Education, School of Physics and Astronomy, Beijing Normal University, Beijing 100875, China}
\affiliation{Institute of Radiation Technology, Beijing Academy of Science and Technology, Beijing 100875, China}

\author{F.~L.~Liu}
\affiliation{China Institute of Atomic Energy, Beijing, 102413, China}
\affiliation{Center for Nuclear Study, The University of Tokyo, Wako-shi, 351-0198, Japan}

\author{P.~Luan}
\affiliation{School of Nuclear Science and Technology, University of South China, Hengyang 421001, China}

\author{S.~S.~Lv}
\affiliation{Key Laboratory of Beam Technology of Ministry of Education, School of Physics and Astronomy, Beijing Normal University, Beijing 100875, China}
\affiliation{Institute of Radiation Technology, Beijing Academy of Science and Technology, Beijing 100875, China}

\author{M.~L.~Qiu}
\affiliation{Key Laboratory of Beam Technology of Ministry of Education, School of Physics and Astronomy, Beijing Normal University, Beijing 100875, China}
\affiliation{Beijing Normal University at Zhuhai, Zhuhai 519087, China}
\affiliation{Institute of Radiation Technology, Beijing Academy of Science and Technology, Beijing 100875, China}

\author{S.~K.~Shao}
\affiliation{NUCTECH Company, Beijing 100084, China}

\author{Q.~L.~Shuai}
\affiliation{Beijing Normal University at Zhuhai, Zhuhai 519087, China}
\affiliation{Key Laboratory of Beam Technology of Ministry of Education, School of Physics and Astronomy, Beijing Normal University, Beijing 100875, China}

\author{Q.~Tang}
\affiliation{School of Nuclear Science and Technology, University of South China, Hengyang 421001, China}

\author{H.~R.~Wang}
\affiliation{Key Laboratory of Particle and Radiation Imaging (Ministry of Education) and Department of Engineering Physics, Tsinghua University, Beijing 100084, China}
\affiliation{NUCTECH Company, Beijing 100084, China}

\author{D.~Wu}
\affiliation{State Key Laboratory of Nuclear Physics and Technology, School of Physics, CAPT, Peking University, Beijing 100871, China}

\author{M.~Y.~Wu}
\affiliation{Institute of Advanced Science Facilities, Shenzhen 518107, China}

\author{J.~S.~Xie}
\affiliation{School of Nuclear Science and Technology, University of South China, Hengyang 421001, China}

\author{Z.~H.~Zhang}
\affiliation{China Nuclear Power Technology Research Institute Co., Ltd., Shenzhen, 518000, China}

\author{Z.~H.~Zhang}
\email{Corresponding author: zhenhua@bnu.edu.cn}
\affiliation{Beijing Normal University at Zhuhai, Zhuhai 519087, China}
\affiliation{Key Laboratory of Beam Technology of Ministry of Education, School of Physics and Astronomy, Beijing Normal University, Beijing 100875, China}

\date{\today}

\begin{abstract}
Despite compelling observational evidence for dark matter (DM), its fundamental physical properties remain poorly understood. In this report, we propose a three-step light-shining-through-walls (LSW) experimental scheme utilizing the high-brilliance, high-energy X-rays from the ID21 Hard X-ray Imaging Beamline at the High Energy Photon Source (HEPS) to search for signatures of dark photons (DPs) and other weakly interacting slim particles (WISPs). The scheme includes three steps of LSW experiments: a short-term (several days) dedicated exposure experiment, a long-term (several years) synchronous accompanying experiment, and a WISP detection with strong magnetic fields. Projection results show that this HEPS-based LSW experiment can effectively constrain DP parameters in the 1 eV--400 keV mass range, covering unexploited parameter space of the existing X-ray LSW experiments. It provides a least model-dependent and most purely-laboratory approach for probing dark sector particles and advancing new physics research beyond the Standard Model gradually.

\end{abstract}

\maketitle

\section{Introduction}
Nearly a century has passed since F. Zwicky published the first observational evidence for the existence of dark matter (DM) in 1933~\cite{Zwicky1933}. Despite abundant compelling observational evidence confirming its presence, humans still understand very little about the physical properties of DM, which accounts for 26.4\% of the universe's critical density (equivalent to 84.4\% of the total matter density)~\cite{PartPhysRev2024}.

The Standard Model (SM) has achieved great success in explaining and predicting almost all the experimental results. However, some observational evidence, such as DM~\cite{DMProblem_1987} and neutrino oscillations~\cite{NeutrinoOscillation_1998, NeutrinoOscillation_2002}, shows that the SM must be extended. Numerous theories beyond the Standard Model (BSM) have been proposed to describe DM. In parallel, a wide array of DM experiments are underway to verify these novel DM models. Especially after the survival parameter space for WIMPs has been pushed close to the neutrino floor, researchers have gradually broadened or shifted their attention to a far broader range of DM candidates. We prefer to categorize DM experiments on the basis of the locations of their detectors: underground, ground-based, and space-borne observatories, with space missions largely corresponding to indirect detection. Unfortunately, to date, none of these experiments have detected a definite DM signal, although several suspicious excesses have been observed in the past. Alternatively, we may adopt an experimental exploration perspective: under purely-laboratory conditions, priority should be given to examining minimal extensions of the SM toward the dark sector that are fully model-independent on other models. One of the most representative experiments is the light-shining-through-walls (LSW) experiment.

Photons cannot penetrate walls because of their large interaction cross sections with ordinary matter. However, if photons can convert into DM particles, they can pass through the walls. After traversing the barrier, these particles would convert back into photons, reproducing the apparent effect of light shining straight through the walls. Starting from dark photon (DP, denoted as ``$\gamma'$''; also known as hidden photon or paraphoton), this report presents a three-step proposal for conducting LSW experiments at the High Energy Photon Source (HEPS):

(1) Conduct a short-duration exposure experiment (on the timescale of days) to search for photon--DP oscillation signatures;

(2) Carry out an accompanying experiment capable of long-duration exposure (on the timescale of months or years);

(3) Deploy strong magnetic fields to search for additional extremely weakly interacting slim particles (WISPs), such as axions, axion-like particles (ALPs), and minicharged particles.

\section{Dark photon sources}
As the gauge boson of a new group of dark sector $\mathrm{U(1)_{DS}}$, DP is a minimal extension of the SM connecting the SM and the dark sector. As shown in Eq.~\ref{eq::Lagrangian}, there is a kinetic mixing parameter of $\chi$ in the Lagrangian that describes the dark sector group $\mathrm{U(1)_{DS}}$ and SM electromagnetic group $\mathrm{U(1)_{QED}}$ at low energies~\cite{Holdom_1986, Foot_1991}. This makes DP difficult but not impossible to observe because of its small $\chi$. Existing constraints on DP are compiled by Cajohare on the website~\cite{AxionLimits, DPhandbook}. We classify the DP sources into three categories: dark-photon dark matter (DPDM), natural sources and artificial sources.

\begin{equation}
\begin{aligned}
\mathcal{L} = -\frac{1}{4}F_{\mu\nu}^2
-\frac{1}{4}B_{\mu\nu}^2
-\frac{1}{2}\chi F^{\mu\nu}B_{\mu\nu}+\frac{1}{2}m_\mathrm{DP}^2B_{\mu}^2,
\label{eq::Lagrangian}
\end{aligned}
\end{equation}
where $F_{\mu\nu}$ is the field strength tensor for the $\mathrm{U(1)_{QED}}$ gauge field $A^{\mu}$, $B^{\mu\nu}$ is the field strength for the $\mathrm{U(1)_{DS}}$ field $B^{\mu}$, and $m_\mathrm{DP}$ may arise via a Higgs or St\"{u}eckelberg mechanism.

i. DPDM

In addition to being a mediator of the SM and dark sector, DP can also be among the DM candidates and is known as DPDM. Under the assumption that DPDM contributes to all local DM density, the DM experiments continuously update the excluded parameter space on $\left ( m_\mathrm{DP}, \chi \right )$.

The electromagnetic detectors search for the DPDM signals in the wavelength bands in which they operate, including SHUKET~\cite{SHUKET_2019}, WISPDMX~\cite{WISPDMX_2019}, HAYSTAC~\cite{HAYSTAC_2021}, Dark $E$ Field~\cite{DarkEfield_2021}, Superconducting Qubit~\cite{Qubit_2021}, ADMX~\cite{ADMX_2022}, FAST~\cite{FAST_2023},   QUALIPHIDE~\cite{QUALIPHIDE_2023}, DOSUE-RR~\cite{DOSUE-RR_2023}, CAPP~\cite{CAPPMAX_2024}, SHANHE~\cite{SHANHE_2024}, and SQMS~\cite{SQMS_2024}. The ionizing radiation detectors impose the constraints on $\chi$ as helioscopes, including DAMIC~\cite{DAMIC_2017, DAMIC_2019}, MAJORANA~\cite{MAJORANA_2017, MAJORANA_2024}, XMASS~\cite{XMASS_2018}, EDELWEISS~\cite{EDELWEISS_2018}, SENSEI~\cite{SENSEI_2025}, SuperCDMS~\cite{SuperCDMS_2019, SuperCDMS_2020}, CDEX~\cite{CDEX_2020}, GERDA~\cite{GERDA_2022}, and XENON~\cite{An_2013, XENON1T_ER2020, XENONnT_2022}.

ii. Natural sources

Because kinetic mixing occurs between DP and the SM photon, any photon source, such as the Sun and cosmic microwave background (CMB),  can clearly be regarded as a DP source. From the perspective of the Earth, the Sun is the most intense DP source, with a flux reaching $\sim \left( 10^{35} \cdot \chi^2 \right)$ s$^{-1}\cdot$ cm$^{-2}\cdot$ eV$^{-1}$ arriving on the Earth. Solar DPs are searched by CAST~\cite{CAST_2008}, LOFAR~\cite{An_2024} and some of the detectors mentioned above, such as MAJORANA~\cite{MAJORANA_2017}, XENON~\cite{SDP_XENON1T}, and CDEX~\cite{CDEX_2020}.

Moreover, if DPs exist, then some natural photon sources would cool or decay more rapidly than we can observe with ordinary photons. We consider these experiments indirect detections. For example, some constraints on DP are given on the basis of the CMB distortion~\cite{Fixsen_1996, Caputo_2020, Caputo_2020_PRL, Alessandro_2009, Garcia_2020}, stellar cooling~\cite{Redondo_2013, Vinyoles_2015, Hong_2021}, and black-hole spin-down~\cite{Stott_2020, Ghosh_2021, Aswathi_2025}. Studies based on these natural sources are inevitably model-dependent on astrophysics and astronomy.

iii. Artificial  sources

\begin{figure}[!htbp]
\centering 	
\includegraphics[width=0.48\textwidth]{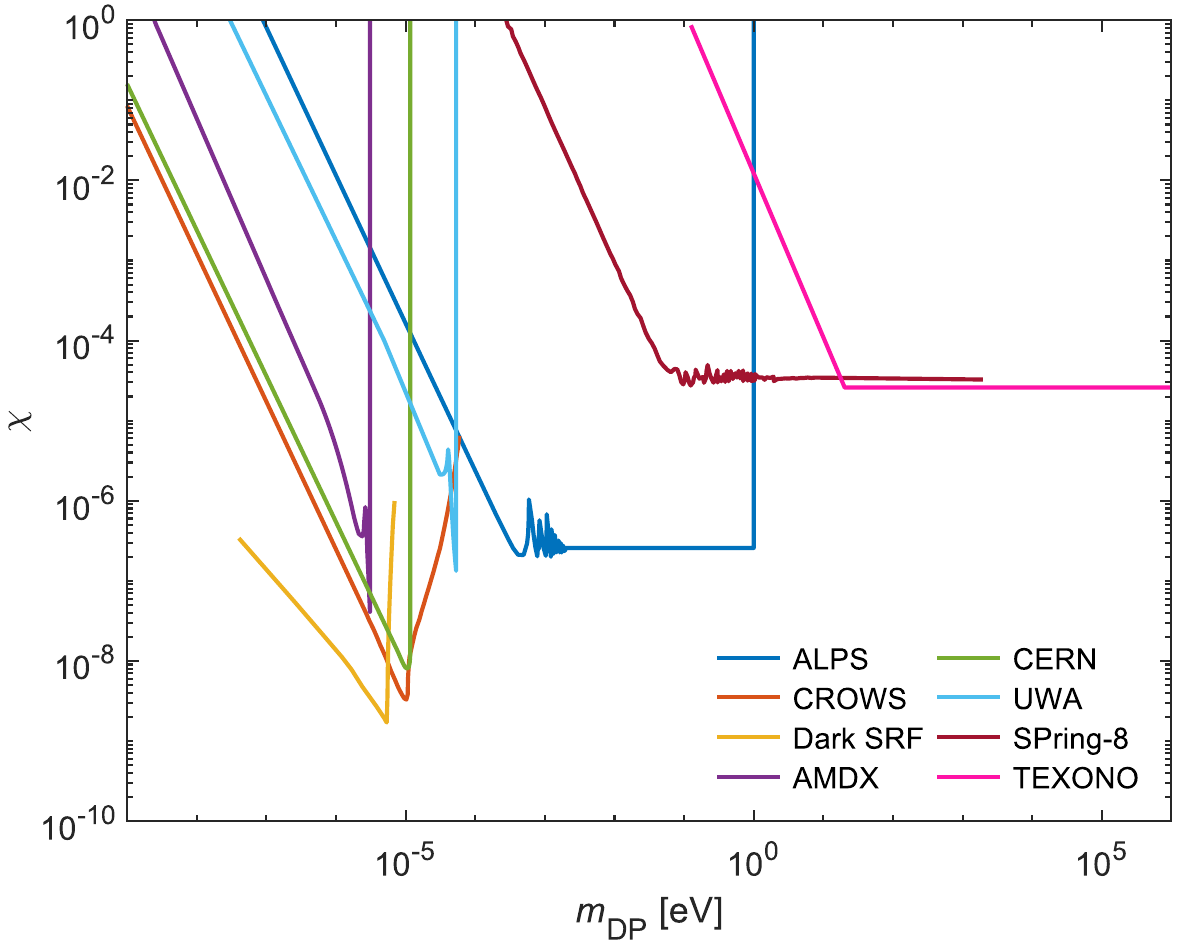}      
\caption{Constraints from LSW experiments searching for DPs from artificial sources~\cite{ALPS_2010, CROWS_2013, DarkSRF_2023, AMDX_2013, CERN_2012, SPring8_2013, UWA_2010, Reactor_2019}.}
\label{fig::ArtLSWResult}
\end{figure}

Nuclear reactors~\cite{Reactor_2017, Reactor_2019, Reactor_2021}, fixed targets with accelerators~\cite{Gninenko_2014, DarkSHINE_2022}, colliders~\cite{NA64_2019, Zhang_2019, LDMX_2023, BESIII_2023, BESIII_2024}, X/$\gamma$-ray sources~\cite{SPring8_2013}, and optical laser devices~\cite{Ahlers_2008, BMV_2008, GammeV_2008, LIPSS_2008, ALPS_2022} are used as artificial DP sources. The least model-dependent and the most purely-laboratory bounds are from the LSW experiments in the photon--DP oscillation mode with artificial sources. These are verifiable results because suspicious signals can be confirmed by switching the photon sources on and off. However, the existing constraints~\cite{ALPS_2010, CROWS_2013, DarkSRF_2023, AMDX_2013, CERN_2012, SPring8_2013, UWA_2010, Reactor_2019}, as shown in Fig.~\ref{fig::ArtLSWResult}, are relatively weak at $m_\mathrm{DP}>1$ eV.

In the optical band, the light intensity can be multiplied many times through multiple reflections. However, in the X/$\gamma$-ray band, it is impossible to increase the equivalent photon intensity along the transport path; only the original photon intensity can be enhanced. A nuclear reactor can be considered a DP source with high energy and large flux. However, because the number of $\gamma$ photons in the reactor core cannot be directly measured, this introduces some numerical model dependence. Furthermore, from the perspectives of safety and cost efficiency, significantly increasing the number of photons in the reactor is difficult. Fixed targets with accelerators and colliders are typically set to cover the region of 1 MeV $\leq m_\mathrm{DP} \lesssim$ 10 GeV based on the escape detection of DPs, which are considered indirect detection experiments. Because the radioisotope source emits photons isotopically, its activity needs to be much greater than $10^{13}$ Bq to achieve a competitive physical result. This is a significant challenge for the radiation protection of personnel. Currently, the only possible option seems to be the X-ray source.

X-ray sources have undergone more than 120 years of development, and their brilliance, measured in photons $\cdot$ s$^{-1}\cdot$ mm$^{-2}\cdot$ mrad$^{-2}\cdot$ 0.1\%BW$^{-1}$, has increased significantly from $10^{7}$ to $10^{24}$. Although their flux is much lower than that of the Sun, their higher X-ray energy ($\omega \gtrsim 100$ keV) and the reduced model dependence motivate a re-examination of LSW experiments using X-ray sources.

\section{LSW at the HEPS}

\begin{figure}[!htbp]
\centering 	
\includegraphics[width=0.48\textwidth]{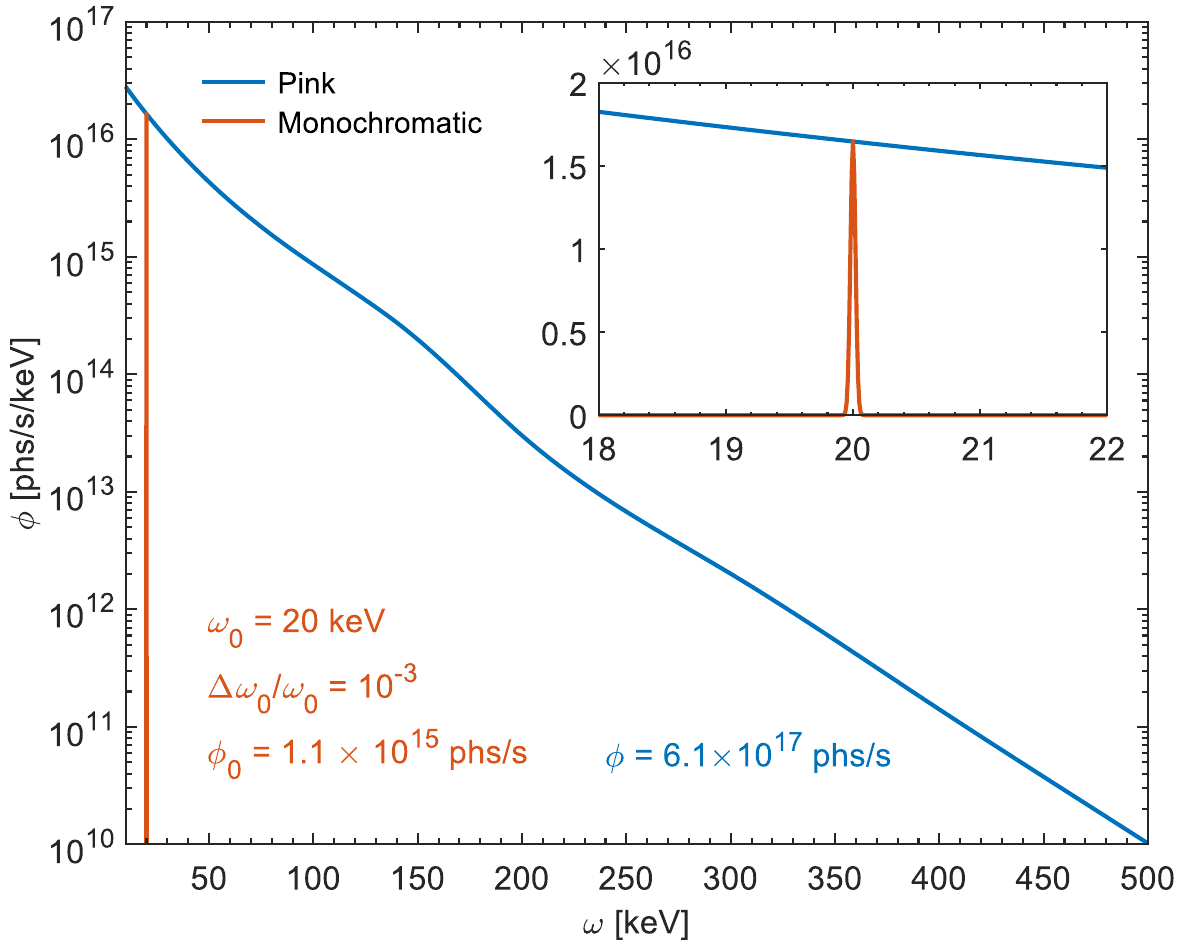}      
\caption{An example of the X-ray photon flux spectrum of the ID21-High Energy X-ray Imaging Beamline at the HEPS (31138.02.HEPS.ID21). The blue line represents the example flux when one of the photon insertion devices (PIDs) is used; the orange line represents the example flux when the monochromator is used, with a relative energy bandwidth of $\Delta \omega/ \omega = 10^{-3}$ at 20 keV, and the passing ratio can be approximately 1 in the pass-band.}
\label{fig::XraySpec}
\end{figure}

The HEPS, as the fourth-generation synchrotron radiation photon facility with the highest spectral brilliance in the world, started trial operations and user experiments in December 2025. An example of the X-ray photon flux spectrum of the ID21-High Energy X-ray Imaging Beamline at the HEPS (31138.02.HEPS.ID21, denoted as ID21) is shown in Fig.~\ref{fig::XraySpec}. The beamline spot size is on the order of millimeters, while the detector size is on the order of centimeters. Therefore, in this report, the dimension of the photon flux of $\phi$ omits the area and directly uses the unit of photons/s/keV, or it can be referred to as ``differential photon number".

ID21 is equipped with three types of photon insertion devices (PIDs): a cryogenic permanent-magnet undulator, an in-air wiggler, and a mango pendulum. The flux of synchrotron radiation photons generated by electrons in them is similar. Using all three PIDs simultaneously can achieve three times the photon intensity. The pink synchrotron light from the PID is filtered into monochromatic light by a monochromator with a relative energy bandwidth of $\Delta \omega/ \omega = 10^{-4}$--$10^{-2}$. The photon loss within the pass-band is small; thus, the passing ratio can be approximately regarded as 1 here. However, all the photons outside the pass-band are attenuated, resulting in a significant loss of photon flux (for example, as shown in Fig.~\ref{fig::XraySpec}, decreasing sharply from $6.1\times10^{17}$ to $1.1\times10^{15}$ photons/s).

Analogous to neutrino oscillations, SM photons and DPs oscillate back and forth into one another~\cite{Okun_1982}. The well-known probability for one quantum oscillation at a travel distance of $L$ in a vacuum is given in Eq.~\ref{eq::Pocs}. 

\begin{equation}
\begin{aligned}
    P_{\gamma \leftrightarrow \gamma'} \simeq 4 \chi^2 \cdot \sin^2
    \left( \frac{\Delta k\cdot L}{2} \right),
    \label{eq::Pocs}
\end{aligned}
\end{equation}
where $\omega$ is the energy of the SM photons and $\Delta k= \omega-\sqrt{\omega^2-m_\mathrm{DP}^2}$ is the difference in momentum.

\begin{figure}[!hbtp]
\centering 	
\includegraphics[width=0.48\textwidth]{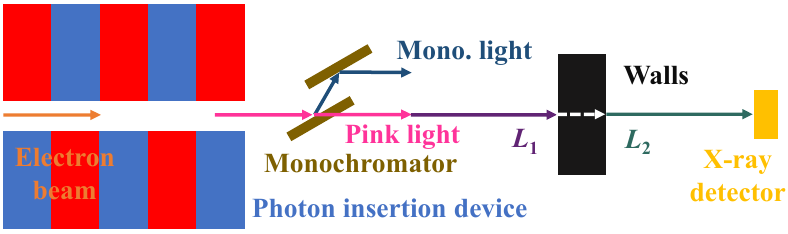}      
\caption{LSW experimental setup in the X-ray beamline.}
\label{fig::LSWSetup}
\end{figure}

As shown in the experimental setup in Fig.~\ref{fig::LSWSetup}, there are four major physical processes (including two oscillations, $\gamma \rightarrow \gamma'$ and $\gamma' \rightarrow \gamma$) in the LSW experiments: (i) the photon source emits X-ray photons; (ii) X-ray photons may oscillate into DPs in the vacuum $L_1$ and then travel through walls of ordinary matter; (iii) DPs may oscillate back into X-ray photons in the vacuum $L_2$; and (iv) the X-ray detector behind walls may catch the X-ray photons and give physical signals on DPs. Standing behind the walls, we may see the photons reproduced with the photon--DP oscillations, similar to the light shining through the walls. Considering the photon detection efficiency ($\eta<1$) of the detector, the number of regenerated photons detected per unit time behind the walls is shown in Eq.~\ref{eq::ndet}.

\begin{equation}
\begin{aligned}
    n_\mathrm{det}(\omega) & = n_\mathrm{s} \cdot \eta =  n_{\gamma} \cdot P_{\gamma \rightarrow \gamma' \rightarrow \gamma} \cdot \eta \\
    & \simeq n_{\gamma} \cdot 16\chi^4 \sin^2\left( \frac{\Delta k\cdot L_1}{2} \right)
    \sin^2\left( \frac{\Delta k\cdot L_2}{2} \right)\cdot \eta,
    \label{eq::ndet}
\end{aligned}
\end{equation}
where $n_{\gamma}$ is the differential photon number in the unit of photons/s/keV and is regarded as the photon flux of $\phi$ at the starting point; $L_1$ is the path length between the starting point and the wall; and $L_2$ is the distance between the wall and the detector. Because $\chi \ll 1$, the attenuation of DPs in the walls can be neglected. The DP signal of $n_\mathrm{s}$ can be expressed by Eq.~\ref{eq::ns}.

\begin{equation}
\begin{aligned}
    n_\mathrm{s}\pm{\sigma_\mathrm{s}} = (n_\mathrm{on} - n_\mathrm{off}) \pm \sqrt{\frac{n_\mathrm{on}}{t_\mathrm{on}}+\frac{n_\mathrm{off}}{t_\mathrm{off}}},
    \label{eq::ns}
\end{aligned}
\end{equation}
where $n_\mathrm{on}$ ($t_\mathrm{on}$) and $n_\mathrm{off}$ ($t_\mathrm{off}$) represent the count rates (running time) when the X-ray beamline is on and off, respectively. Before a definite signal is detected, we adopt the 0-signal assumption; that is, $n_\mathrm{s} = 0$ and $n_\mathrm{on} = n_\mathrm{off} = n_\mathrm{bkg}$. The uncertainty of the DP signal is subsequently determined by the background and the running time. For convenience, let $t = t_\mathrm{on} = t_\mathrm{off}$ in the following discussion.

\section{Radiation background}
\begin{figure}[!htbp]
\centering 	
\includegraphics[width=0.48\textwidth]{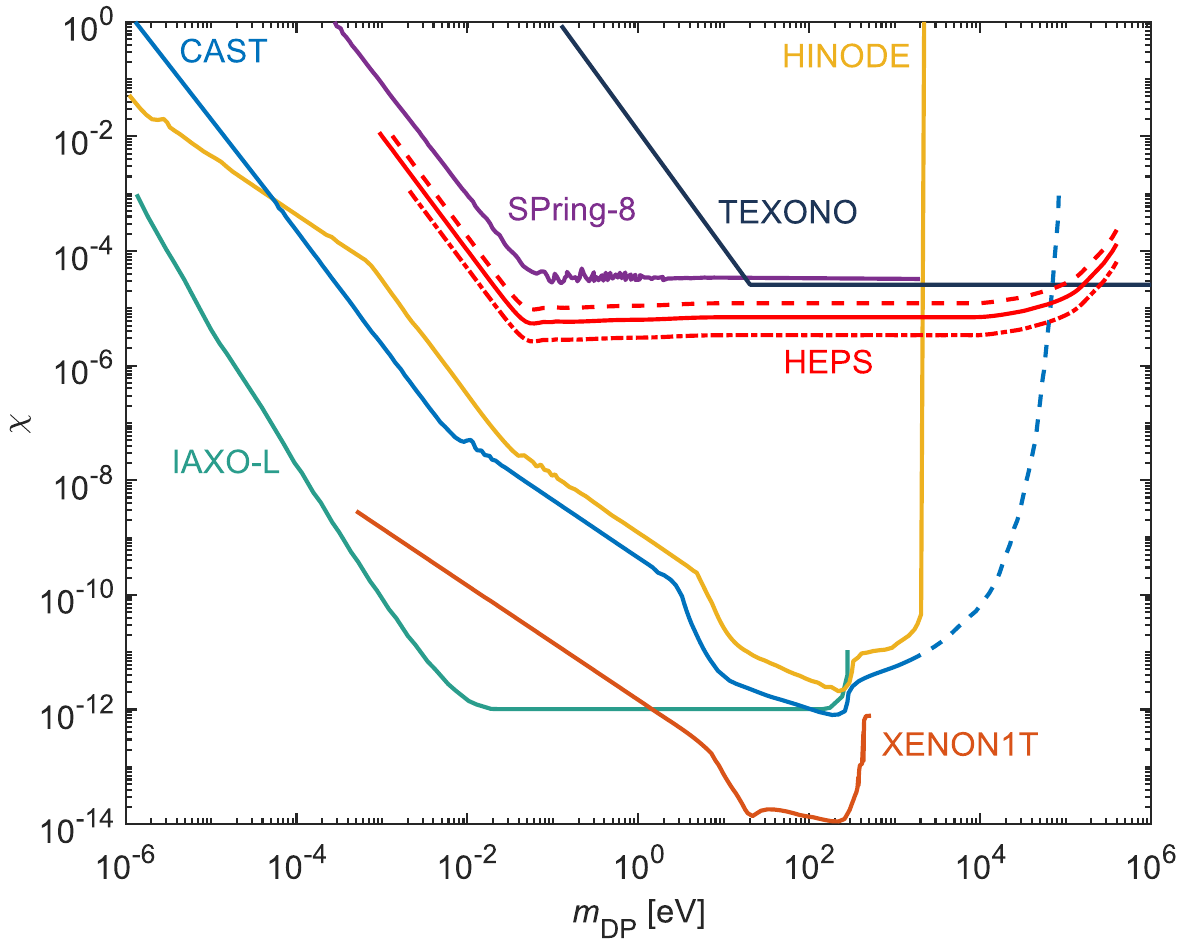}
\caption{Constraints from LSW experiments in X/$\gamma$-ray band searching for DPs
from the Sun~\cite{CAST_2008, HINODE, IAXO, SDP_XENON1T}, X-ray beamline~\cite{SPring8_2013} and nuclear reactor~\cite{Reactor_2019}. The results from CAST are represented by a dotted line when $m>15$ keV, because the limits are continued by assuming the detector is equally sensitive and has no threshold~\cite{CAST_2008}.}
\label{fig::LSWResults}
\end{figure}

To maximize experimental sensitivity, DM detectors are preferably deployed in deep underground laboratories, where mountain rock and crustal strata attenuate the radiation backgrounds induced both directly and indirectly by cosmic rays (CRs). As shown in Fig.~\ref{fig::LSWResults}, CAST~\cite{CAST_2008}, HINODE~\cite{HINODE}, IAXO-L~\cite{IAXO} and XENON1T~\cite{SDP_XENON1T} were used to search for solar DPs. XENON1T achieved an unparalleled level of sensitivity ($\chi \sim 10^{-14}$ at $m_\mathrm{DP} \sim 100$ eV) because of its extremely low background rate of 76 counts/tonne/year/keV between 1 and 30 keV and very large exposure of 0.65 tonne-years~\cite{XENON1T_ER2020} at the Gran Sasso National Laboratory, located 1400 m underground.

\begin{figure}[!htbp]
\centering 	
\includegraphics[width=0.48\textwidth]{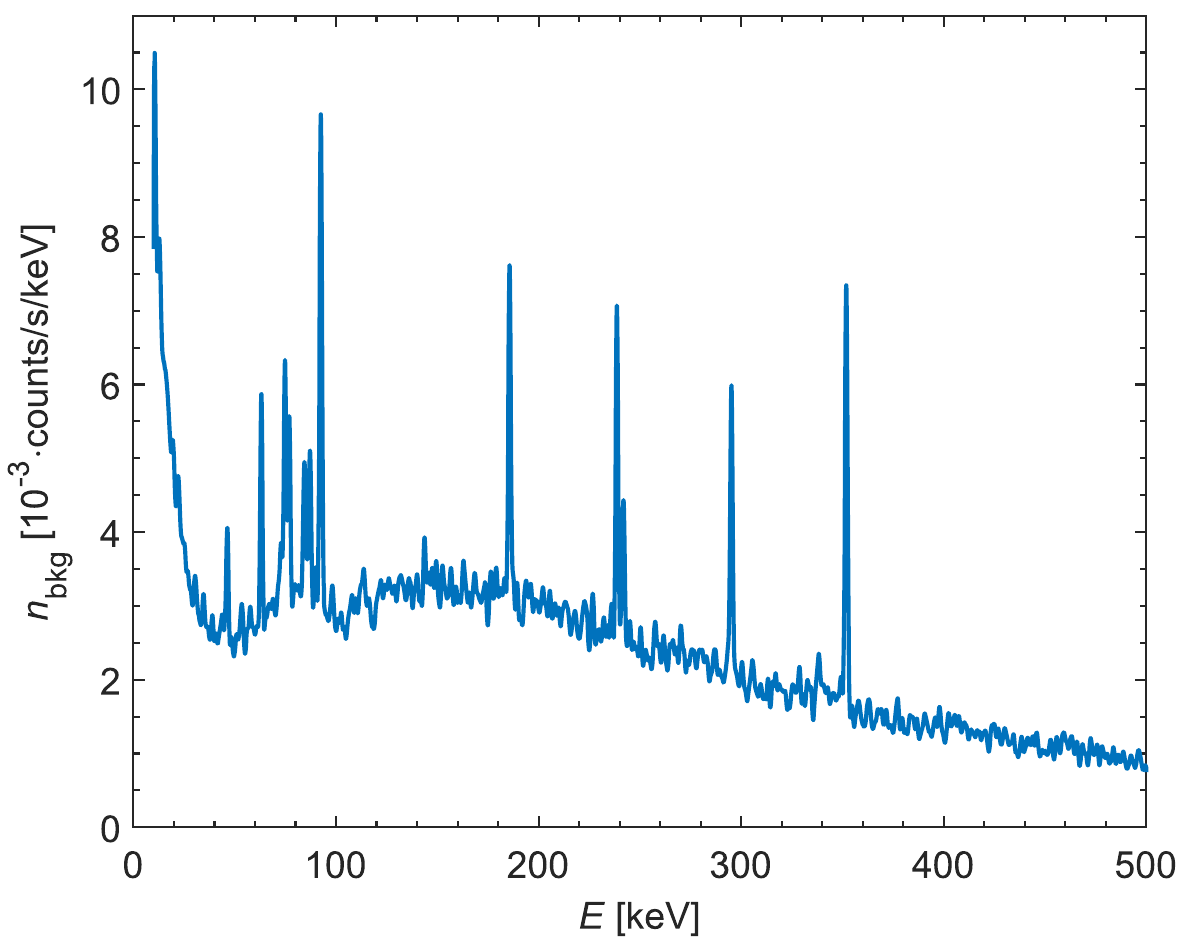}
\caption{A radiation background spectrum measured by an HPGe detector (with a mass of 989 g) within a lead chamber in Beijing.}
\label{fig::bkg}
\end{figure}

\begin{figure}[!htbp]
\centering 	
\includegraphics[width=0.48\textwidth]{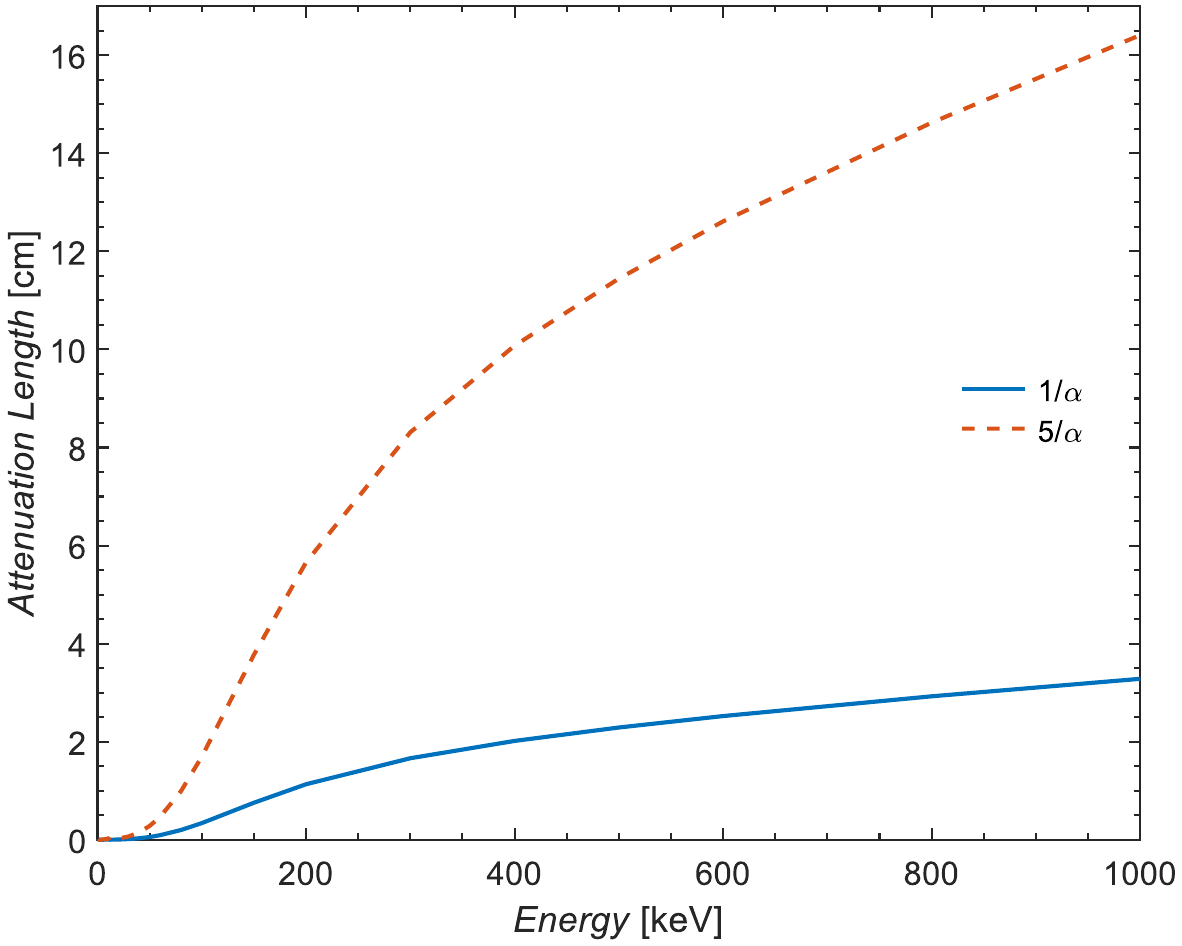}
\caption{Attenuation length (1/$\alpha$) of X-rays in germanium~\cite{XrayAttenuCoff}.}
\label{fig::AttenuationLength}
\end{figure}

A background spectrum measured by an HPGe detector (with a crystal thickness of 6 cm, a mass of 989 g) within a lead chamber (with an outer layer of lead that is 10 cm thick and an inner layer of copper that is approximately 3 mm thick) in Beijing is shown in Fig.~\ref{fig::bkg}. The current background level is comparable to that of the HPGe detector (with a crystal thickness of 2.5 cm and a mass of 370 g) used by T. Inada et al. at SPring-8~\cite{SPring8_2013}, although direct comparison between different detectors is not advisable here. For instance, in rare-event experiments, the background is usually described in units of counts/keV/s/kg~\cite{MaJingLu}. Obviously, using detectors with smaller volumes (or masses) will result in a lower background level in terms of counts/keV/s. The attenuation length ($1/\alpha$) of X-rays in germanium is shown in Fig.~\ref{fig::AttenuationLength}. The X-ray energy of ID21 can reach up to approximately 500 keV at its maximum. Thus, a thinner HPGe crystal is not very suitable for X-ray detection at the 100 keV level.

Locating the hit positions in HPGe based on pulse shape analysis (PSA)~\cite{Dai_PSD} might be the best solution to this contradiction. The hit position can be inferred on the basis of the pulse shape collected by the data acquisition (DAQ) system. In the radial direction, if the hit position deviates from the projection position of the X-ray spot, then this event can be determined as a background event and excluded from the dataset; in the axial direction, combined with the measured energy, if the hit position deviates from the penetration distance of the X-ray of that energy, then this event will be excluded. This effectively achieves the geometric segmentation of HPGe at the algorithmic level.

\begin{figure}[!htbp]
\centering 	
\includegraphics[width=0.48\textwidth]{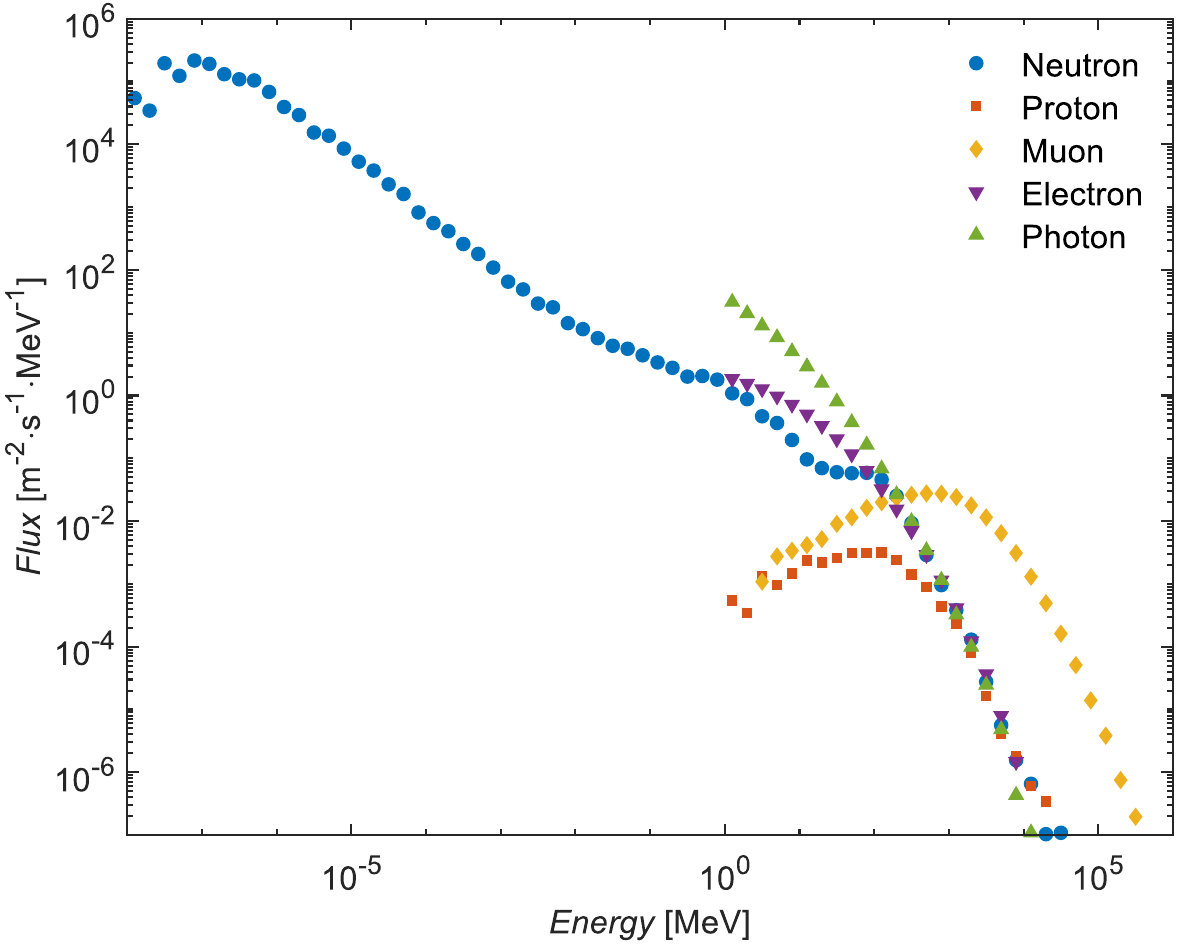}
\caption{Flux spectra of the components of cosmic rays (CRs) at sea level in Beijing calculated by $\tt Cosmic-ray\quad Shower\quad Library\tt$.}
\label{fig::CRFlux}
\end{figure}

The fluxes of the CR components at sea level in Beijing can be estimated by $\tt Cosmic-ray\quad Shower\quad Library\tt$, and the spectra are plotted in Fig.~\ref{fig::CRFlux}. Neutrons and high-energy muons contribute mainly to the cosmogenic background and CR background respectively. High-energy muon backgrounds can be rejected via active cosmic-ray anticoincidence (CRAC) detectors. With the detector mounted horizontally, CRAC and Compton-scattering anticoincidence (CSAC) systems can be integrated into a single module. Neutrons, however, require a mixed shielding structure: heavy nuclei (such as iron) are used to attenuate fast neutrons, and then light nuclei (such as water and polyethylene) are used to slow the neutrons (doping with boron can absorb some thermal neutrons). The ceiling of the building can also help to achieve some CR attenuation by chance. With a classical lead-OFHC (oxygen-free high-purity copper) structure, the X/$\gamma$ radiation background can be further reduced. We aim to reduce the experimental background to one order of magnitude lower than the current level by an active and passive hybrid background shielding system.

\section{Data acquisition}
\begin{figure}[!htbp]
\centering 	
\includegraphics[width=0.48\textwidth]{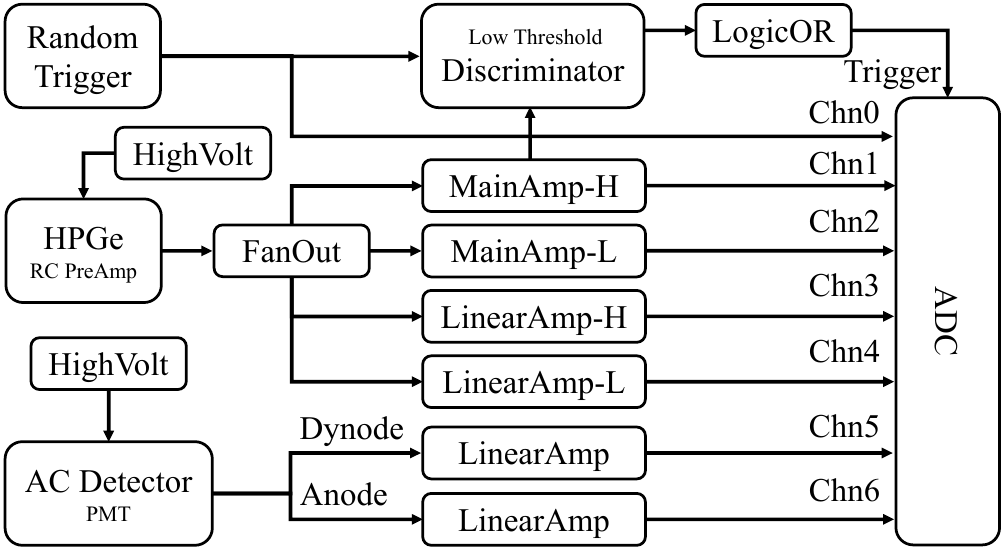}
\caption{Block diagram of the data acquisition (DAQ) system setup.}
\label{fig::DAQSetup}
\end{figure}

On the basis of the requirements of PSA, the DAQ system is inclined toward more digitization, while appropriate analog modules are retained to facilitate monitoring of the system status, as shown in Fig.~\ref{fig::DAQSetup}. HPGe is the central detector. The signal from the resistor--capacitor feedback preamplifier is fanned out into four paths: two of them enter the main amplifiers with high and low gains for spectroscopy, and the other two enter the linear amplifiers with high and low gains to meet the ENOB (effective number of bits) requirements for PSA in different energy ranges. The dynode signal and anode signal from the anti-coincidence (AC) detector for CSAC and CRAC are directly input to analog-to-digital conversion (ADC) for subsequent PSA processing. The random trigger (RT) is a rectangular pulse signal with a fixed frequency (reference value of 0.05 Hz) produced by a signal generator. The RT signal is a random signal relative to the HPGe signal and is used to monitor the detector's working status and correct the detection efficiency. The ADC is triggered after a Boolean logic OR operation is performed on the HPGe signal and RT signal.

\section{Projection sensitivity}
The beam-on spectrum ($n_\mathrm{on}$) and the beam-off spectrum ($n_\mathrm{off}$) were compared via minimum-$\chi^2$ analysis as shown in Eq.~\ref{eq::ChiSquare}. Under the 0-signal assumption, the sensitivity is estimated from the background. The 90\% confidence level (CL) limit is estimated with $\Delta \chi^2_\mathrm{anal} = 2.71$, which is based on the Feldman-Cousins unified approach~\cite{FCChiSquare}.

\begin{equation}
\chi^2_\mathrm{anal} = \sum_j \frac{\left(n_{\mathrm{on,}j}-n_{\mathrm{off,}j}\right)^2}
{(\Delta_{\mathrm{on,}j})^2+(\Delta_{\mathrm{off,}j})^2}
\simeq  \sum_j \frac{\left(n_{\mathrm{s,}j}\right)^2}
{2\cdot(\Delta_{\mathrm{bkg,}j})^2},
\label{eq::ChiSquare}
\end{equation}
where $j$ represents the $j$-th bin in the spectra.

In the first step of the proposal, we have two options. One is the baseline plan, Plan A, which conducts the experiment using all the parameters that have already been achieved. Another feasible, but challenging, plan that still requires further effort is to use all three PIDs simultaneously while reducing the background by one order of magnitude. This configuration is denoted as Plan B. When the running time is $t = 10^5$ s, the projection sensitivities of Plan A and Plan B are plotted with dashed lines and solid lines, respectively, in Fig.~\ref{fig::LSWResults}. Here, we set $L_1 = L_2 = 5 $ m. The shed ID21 can support a $(L_1 + L_2)$ of up to 330 m, whereas a longer $L$ can support exploration for lower $m_\mathrm{DP}$.

In the SPring-8 LSW experiment, the DP flux was $\mathcal{O}(10^{13} \cdot \chi^2)$ photons/s/keV, the running time was $\mathcal{O}(10^5)$ s, and the background was not significantly different from that shown in Fig.~\ref{fig::bkg}. Although the DP flux in the reactor core is as high as $\mathcal{O}(10^{21} \cdot \chi^2)$ photons/s/keV, owing to the isotropic emission and the 28 m source--detector distance, the DP flux at the detector is only $\mathcal{O}(10^{13} \cdot \chi^2)$ photons/s/keV. With a larger lower-background exposure dataset ($414\pm100.6$ counts in 160 days~\cite{TEXONO2010}), TEXONO outperforms SPring-8 and expands the coverage to the MeV mass range.

As shown by the red dotted line in Fig.~\ref{fig::LSWResults}, using the pink light from ID21 can achieve a competitive result with the least model-dependence in Plan A. If Plan B is carried out, the photon flux will increase by a factor of 3, the background will decrease by one order of magnitude, and the limit will be updated at $m > 67$ keV. If the running time is extended from 1 day to 1 year, then the solid line in Fig.~\ref{fig::LSWResults} will move down to the position indicated by the dotted line, and the projection sensitivity value for $\chi$ will be reduced to 1/2. From Eq.~\ref{eq::ndet} and \ref{eq::ns}, it can be concluded that $\chi^4 \propto \sqrt{n_\mathrm{bkg}/t}$. $t$ is extended by 315 times, and then, the sensitivity of $\chi$ is improved by only $\sqrt[8]{1/315} = 0.5$.

Dedicating continuous long-term beamtime solely to extend exposure, or relocating the beamline underground to suppress cosmic-ray backgrounds, would impose excessive resource costs for improvements in experimental sensitivity. Perhaps we can envision a dedicated LSW experimental beamline in the future. But this is not feasible at present. Therefore, we should consider the second step in the proposal: conducting an accompanying LSW experiment.

\section{Accompanying LSW experiment}
\begin{figure}[!htbp]
\centering 	
\includegraphics[width=0.45\textwidth]{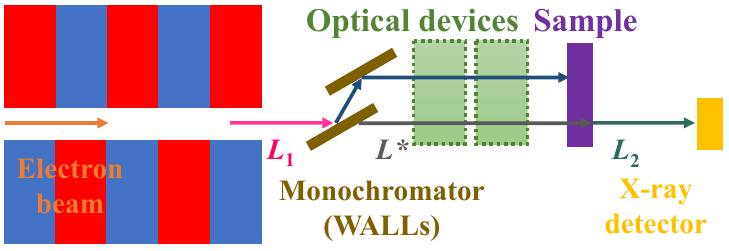}
\caption{Schematic view of the accompanying LSW experiment in the X-ray beamline. See text for details.}
\label{fig::AccExpSetup}
\end{figure}

Synchrotron radiation X-ray beamlines are the large-scale user facilities for photon science and play important roles in research across physics, chemistry, material, life, energy and environmental sciences. A schematic view of the accompanying LSW experiment newly designed in the proposal is shown in Fig.~\ref{fig::AccExpSetup}. In the main-line experiment, X-rays pass through a double-crystal monochromator and a series of optical devices before being directed onto the sample as a probe. Then, the main-line user infers the properties of the sample on the basis of the behavior of the X-rays within it.

Treating the monochromator as a wall turns the LSW into a byproduct of the main-line experiments in the X-ray beamline. Note that the choice of walls may change in the different main-line experiments. The function of the walls is to block SM photons and only allow DPs to pass through. Any device in the optical path that significantly attenuates the flux of X-ray photons can be considered a candidate for the walls. At this point, the X-ray photon may oscillate into a DP in the $L_1$ section between the PID and the front crystal of the monochromator, successfully penetrating into the $L^*$ section. The distance between the photon beams entering and exiting the monochromator is usually in the range of centimeters. This means that in the $L^*$ section, there are certain optical devices and the main-line user's test sample that exist and act as obstacles for DPs. 

\begin{figure}[!htbp]
\centering 	
\includegraphics[width=0.45\textwidth]{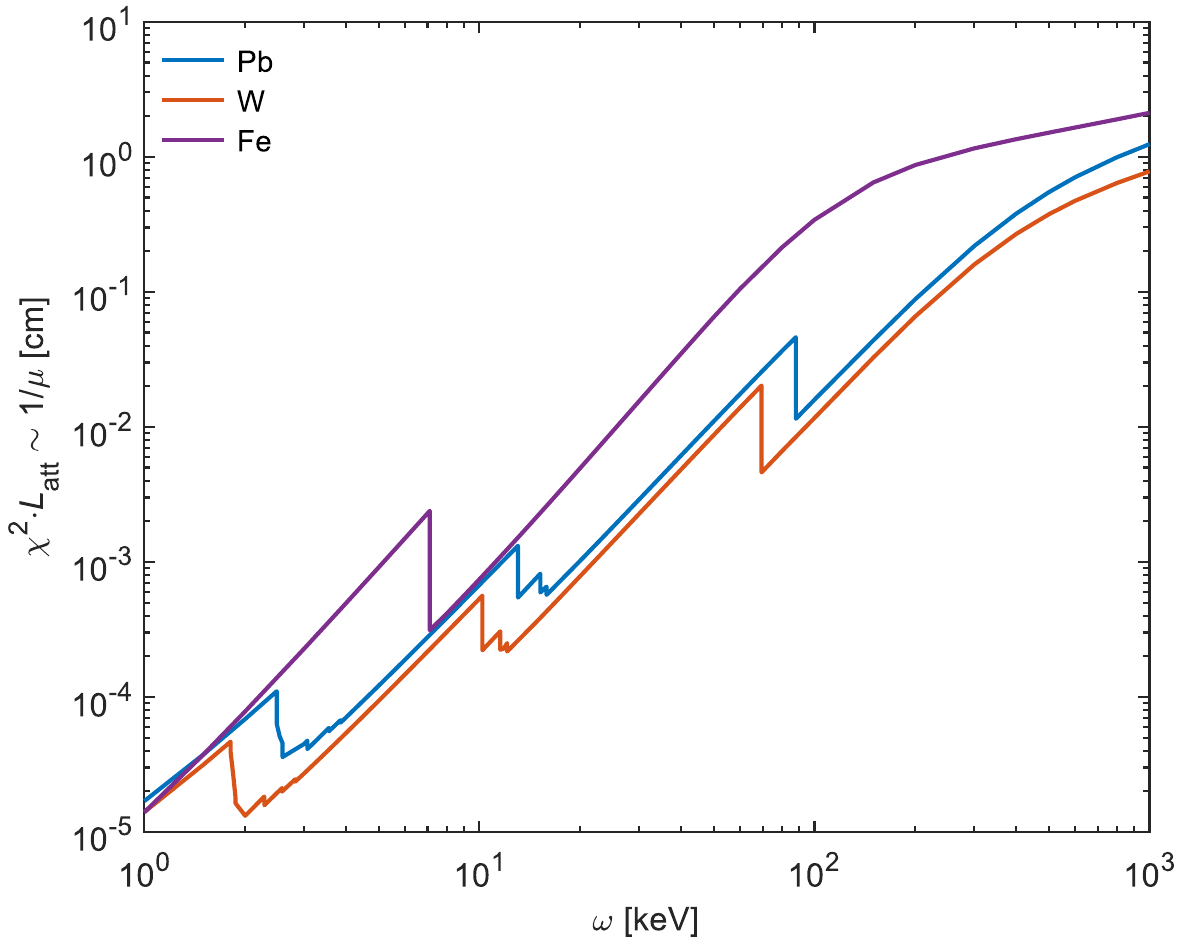}
\caption{The estimated attenuation length of $\chi^2\cdot L_\mathrm{att}$ of DP in presence of the lead (Pb), tungsten (W), and iron (Fe) obstacles.}
\label{fig::DPAttenuation}
\end{figure}

Like the attenuation of X-rays in a medium, we evaluate the attenuation of DPs in a medium by multiplying the linear attenuation coefficient of $\mu$ by $\chi^2$. Set $\chi^2\mu L_\mathrm{att} = 1$, and then $\chi^2 \cdot L_\mathrm{att} \simeq 1/\mu$ can be used to determine the maximum length of obstacles that are allowed in the $L^*$ section when $\chi$ takes a given value. The estimated attenuation lengths $\chi^2\cdot L_\mathrm{att}$ of DPs are plotted in Fig.~\ref{fig::DPAttenuation}. The oscillation attenuation of DPs in tungsten is the most severe. If $\chi = 10^{-3}$, the allowed length of the obstacle is approximately 1 m at $\omega = 10$ keV. This is a sufficiently large value that enables us to conduct the accompanying LSW experiments. After passing through the $L^*$ section, the DP oscillates back into a photon that can be captured by the detector. As a result, LSW can continuously operate along with various main-line experiments.

As an accompanying experiment, we do not impose any requirements on the main-line experiment. This requires that the real-time monitoring of the beamline and the DAQ system must be achieved, as well as the stable and synchronized operation of these systems. The long-term operational stability of HPGe detectors has been verified in various experiments such as reactor neutrino observations (for example, TEXONO). The key point of the realization of this step of the proposal lies in the coordinated operation of the beamline and the detector. In the execution of the first step of the proposal, in addition to pursuing a new physical result, the focus should also be on developing real-time monitoring technologies for beamline parameters (such as the incident photon flux spectrum, X-ray spot size, $L_1$, $L^*$, and $L_2$). There is another technical challenge. When the main-line users scan large-sized samples, the detector should simultaneously track the direction of the X-ray beamline.

\section{Searching for other WISPs}

Axions and ALPs are also the most typical and popular members of the WISP family. The Peccei-Quinn mechanism necessitates the existence of axions, offering a solution to the problem that QCD maintains perfect time-reversal invariance at the precision achievable by modern experiments~\cite{Axion_Peccei, Axion_Wilczek, Axion_Weinberg}. Pseudoscalar bosons that are very light and similar to axions are generically called ALPs; they appear in many motivated SM extensions~\cite{ALP_Andreas, ALP_Jaeckel}. Both of them are also popular candidates for DM, yet experiments have still failed to detect them. The axion and the photon have a coupling vertex in an external homogeneous magnetic field $B$, and they can convert into each other. By replacing $\chi$ in Eq.~\ref{eq::Pocs} with $\frac{g_{\mathrm{a}\gamma}B\omega}{m_\mathrm{a}^2}$ (where $g_{\mathrm{a}\gamma}$ is the axion--photon coupling parameter), the oscillation probability between an axion with a mass of $m_\mathrm{a}$ and a photon can be obtained. The challenge of this step lies in the need to create a strong magnetic field in the $L_1$ section. Unlike the magnetic field in the $L_2$ section, which is not part of the optical path for the main-line users, the impact of the beamline introduction still requires specialized research. 

\section{Summary}

In this report, we have presented a three-step proposal for conducting the LSW experiment at the HEPS:

(1) First step: Search for DPs in a several-day LSW experiment.

Plan A. With $\phi = 6\times10^{17}$ photons/s, $t = 10^5$ s, and $n_\mathrm{bkg}= 3\times10^{-3}$ counts/keV/s at 100 keV, the experimental sensitivity to $\chi$ lies from $1.1\times10^{-5}$ to $2.3\times10^{-4}$, with the DP mass ranging from 1 eV to 400 keV.

Plan B. With $\phi = 2\times10^{18}$ photons/s, $t = 10^5$ s, and $n_\mathrm{bkg} = 3\times10^{-4}$ counts/keV/s at 100 keV, the experimental sensitivity to $\chi$ lies from $6.3\times10^{-6}$ to $1.3\times10^{-4}$, with the DP mass ranging from 1 eV to 400 keV.

(2) Second step: Search for DPs in a several-year accompanying experiment.

Plan C. With $\phi = 2\times10^{18}$ photons/s, $t = 1$ year, $n_\mathrm{bkg} = 3\times10^{-4}$ counts/keV/s at 100 keV, the experimental sensitivity to $\chi$ lies from $3.0\times10^{-6}$ to $6.5\times10^{-5}$, with the DP mass ranging from 1 eV to 400 keV.

(3) Third step - Search for WISPs in the LSW experiments.

Conducting accompanying experiments under stronger magnetic fields can aid in the 
discovery of more WISPs. The technical details still need to be further discussed and studied.

\begin{acknowledgments}
We would like to thank W.~Chao, G.~F.~Chen, Y.~S.~Hao, T.~Inada, G.~Li, Y.~D.~Liu, W.~Y.~Tang, Y.~B.~Wang, L.~Wang, H.~T.~Wong, J.~L.~Yang, and Q.~Yue for valuable discussions and technical support. We also thank the ID21-High Energy X-ray Imaging Beamline  (\url{https://cstr.cn/31138.02.HEPS.ID21}) and the ID42-Test Beamline  (\url{https://cstr.cn/31138.02.HEPS.ID42}) of High Energy Photon Source for providing technical support and assistance in X-ray experimental methods. 
\end{acknowledgments}

\bibliography{DPLSW}

\end{document}